\documentstyle[12pt,
epsfig]{article}
\relax

\def\bes{\begin{eqnarray}}
 \def\ees{\end{eqnarray}}
\def\be{\begin{equation}}
\def\ee{\end{equation}}
\def\bs{\begin{subequations}}
\def\es{\end{subequations}}
\newcommand{\een}{\end{subequations}}
\newcommand{\ben}{\begin{subequations}}
\newcommand{\beq}{\begin{eqalignno}}
\newcommand{\eeq}{\end{eqalignno}}

 \def\kx{\kappa}
 \def\Lx{\Lambda}
 \def\lx{\lambda}

\begin{document}


\begin{center}
{ \Large \bf
Classical Solutions of Higher-Derivative Theories} 
\\
\vspace{1.5cm}
{\Large 
Nikolaos Tetradis 
} 
\\
\vspace{0.5cm}
{\it
Department of Physics, University of Athens,\\
University Campus, Zographou 157 84, Greece
} 
\end{center}
\vspace{3cm}
\abstract{
We present exact classical solutions of the higher-derivative theory that describes 
the dynamics of the position modulus of a probe brane within a five-dimensional bulk. The solutions 
can be interpreted as static or time-dependent throats connecting two parallel branes.
In the nonrelativistic limit the brane action is reduced to that of the Galileon theory. We derive exact solutions
for the Galileon, which reproduce correctly the shape of the throats at large distances, but fail to do so for their central part.
We also determine the parameter range for which the Vainshtein mechanism is reproduced within the brane theory.
}
\newpage

\section{Introduction}

Exact classical solutions of field theories are interesting as they can describe configurations that differ 
substantially from the usual perturbative vacuum.  
The purpose of this letter is to present a class of exact solutions of certain higher-derivative scalar theories in 3+1 dimensions.
The theories have a geometric origin, as they describe hypersurfaces, which we term branes, embedded 
in a higher-dimensional spacetime. From the point of view of the brane observer the action involves
only derivative terms of a particular form for one scalar degree of freedom.

We obtain our solutions by generalizing known static and time-dependent ones for the Dirac-Born-Infeld (DBI) theory with
vanishing gauge fields.
This theory can be viewed as the effective description of a $(3+1)$-dimensional brane embedded in 
a Minkowski bulk spacetime with one additional spatial dimension. 
A known static solution is the catenoidal configuration  
obtained by joining two branches with opposite first derivatives at the point where they
display square-root singularities \cite{callan,gibbons}. 
The result is a smooth surface that looks like a throat or wormhole 
connecting two asymptotically parallel branes. 
A similar time-dependent solution describes two branes connected by a throat whose
radius evolves with time. The throat shrinks to a minimal radius and subsequently
re-expands \cite{dynamical,rizos}. Alternatively, the square-root singularity 
can be viewed as a propagating shock front \cite{heisenberg}. For the expanding configuration, 
energy is transferred from the location of the shock front, where the energy density diverges, 
to the region behind it. 

We consider generalized theories that describe branes embedded 
in a flat bulk spacetime with one additional spatial dimension. The leading contribution to the action is 
given by the volume swept by the brane (the worldvolume), expressed in terms of the induced metric.
It is invariant under arbitrary changes of the brane worldvolume coordinates. We eliminate this gauge freedom by
identifying the brane coordinates with certain bulk coordinates  (static gauge). 
The remaining bulk coordinate becomes a scalar field of the worldvolume theory, with dynamics governed by the
DBI action. More complicated terms can also be included in the effective action. 
They can be expressed in 
terms of geometric quantities, such as the intrinsic and extrinsic curvatures of the hypersurface. In the static gauge
these can be written in terms of the scalar field and its derivatives,
so that we obtain a higher-derivative scalar theory with a particular structure of geometric origin \cite{dbigal,gliozzi}.
Some of the higher-derivative terms can be considered as quantum corrections to the
DBI action \cite{codello}.

The focus of our analysis is on brane theories constructed so that
the equation of motion does not contain field derivatives higher than the second. In this way ghost fields do not appear in the spectrum.  
The most general scalar-tensor theory with this property was constructed a long time ago \cite{horndeski}, 
and rediscovered recenty. It is characterized as the generalized Galileon (see ref. \cite{genegal} and references therein).
A particular example is
provided by the Galileon theory \cite{galileon}, which results from  
the Dvali-Gabadadze-Porrati (DGP) model \cite{dgp} in the decoupling limit.
The connection with the brane picture is made in 
ref. \cite{dbigal}, where it is shown that the Galileon theory can be reproduced 
in the nonrelativistic limit, starting from
the effective action for the position modulus of a probe brane within a (4+1)-dimensional bulk.

\section{The brane and Galileon theories}

We consider the brane theory as formulated in ref. \cite{dbigal}. 
The induced metric on the brane in the static gauge is $g_{\mu\nu}=\eta_{\mu\nu}+\partial_\mu \pi \, \partial_\nu \pi$, where
$\pi$ denotes the extra coordinate of the bulk space. Our convention for the Minkowski metric is $\eta_{\mu \nu}={\rm diag} (-1,1,1,1)$.
The extrinsic curvature is 
$K_{\mu\nu}=-\partial_\mu\partial_\nu\pi/\sqrt{1+(\partial\pi)^2}$. We denote its trace by $K$.
The leading terms in the brane effective action 
are \cite{dbigal}
\begin{eqnarray}
S_\lx&=&-\lx\int d^4x  \sqrt{-g}=-\lx\int d^4 x \sqrt{1+(\partial \pi)^2}
\label{sl} \\
S_K&=&-M^3_5\int d^4x  \sqrt{-g}\, K=M^3_5\int d^4x\, \left([\Pi]-\gamma^2[\phi]\right)
\label{sn} \\
S_R&=&\frac{M^2_4}{2}\int d^4x  \sqrt{-g}\, R=\frac{M^2_4}{2}\int d^4x \, \gamma
\left([\Pi]^2-[\Pi^2] +2\gamma^2([\phi^2]-[\Pi][\phi]) \right),
\label{skb} 
\end{eqnarray}
where $\gamma=1/\sqrt{-g}=1/\sqrt{1+(\partial \pi)^2}$.
We have adopted the notation of ref. \cite{dbigal}, with $\Pi_{\mu\nu}=\partial_\mu\partial_\nu \pi$ and square brackets
representing the trace (with respect to $\eta_{\mu\nu}$) of a tensor. Also, we define
$[\phi^n]\equiv\partial \pi\cdot \Pi^n \cdot \partial\pi$, 
so that $[\phi]= \partial^\mu\pi \, \partial_\mu\partial_\nu\pi\, \partial^\nu\pi$. 
We define the fundamental scale of the theory as $\Lx=\lx^{1/4}$.
We express all dimensionful quantities in units of $\Lx$ in numerical calculations. This is equivalent to setting $\lx=1$. 

The field equation of motion is \cite{dbigal}
\begin{eqnarray}
&&\lx\, \gamma \Bigl\{  [\Pi]-\gamma^2[\phi] \Bigr\}
-M^3_5\gamma^2 \Bigl\{ [\Pi]^2-[\Pi^2]  +2\gamma^2 \left([\phi^2]-[\Pi][\phi] \right)\Bigr\}
\nonumber \\
&&-\frac{M^2_4}{2}\gamma^3 \Bigl\{
[\Pi]^3+2 [\Pi^3]-3[\Pi][\Pi^2]+3\gamma^2\left( 2\left([\Pi][\phi^2]-[\phi^3] \right)
-\left([\Pi]^2-[\Pi^2] \right)[\phi] \right)\Bigr\}=0.
\nonumber \\
&&~
\label{eom} \end{eqnarray}

The Galileon theory \cite{galileon},  which results from  
the DGP model \cite{dgp} in the decoupling limit,
can also be obtained by taking 
the nonrelativistic limit $(\partial \pi)^2 \ll 1$ of the brane theory \cite{dbigal}. In this process,
terms involving second derivatives of the field, such as $\Box \pi$, are not assumed to be small. 
If total derivatives are neglected, the leading terms in the expansion of eqs. (\ref{sl})-(\ref{skb}) give \cite{dbigal}
\be
S^{NR}=\int d^4x\,\Biggl\{-\frac{\lx}{2}  (\partial \pi)^2 +\frac{M^3_5}{2}  (\partial \pi)^2  \Box \pi 
+\frac{M^2_4}{4}  (\partial \pi)^2\left( (\Box \pi)^2  -(\partial_\mu \partial_\nu \pi)^2 \right)
\Biggr\}.
\label{galil} 
\ee
The term of highest order in the Galileon theory, omitted here,  
can be obtained by including in the
brane action the Gibbons-Hawking-York term associated with the Gauss-Bonnet term of $(4+1)$-dimensional gravity. 
The theory can be put in more conventional form  by defining $\lx=\Lx^4$ and
employing the scalar field $\Lx^2 \pi$. As before, we express all quantities in units of $\Lx$ in numerical calculations.
The field equation of motion is
\be
\lx\, [\Pi]
-M^3_5 \left( [\Pi]^2-[\Pi^2] \right)  
-\frac{M^2_4}{2}\left( [\Pi]^3+2 [\Pi^3]-3[\Pi][\Pi^2] \right)=0,
\label{eomnr} \ee
to be compared with eq. (\ref{eom}).
The relation between the brane and Galileon theories implies the existence of solutions similar to the ones we described for
the brane theory.

Before presenting solutions of the equations of motion, we must clarify an important point in their interpretation. 
We shall study configurations of two parallel branes, possibly connected by a throat. 
The form of the term (\ref{sn}), which is reduced to the second term of eq. (\ref{galil}) in the nonrelativistic limit,
breaks the reflection symmetry across an isolated brane.
In a two-brane system, it would affect differently the upper and lower parts of a throat, thus breaking the reflection 
symmetry across the middle plane between the two branes. 
The origin of this counterintuitive behavior can be traced to the DGP model \cite{dgp}. In that construction, the brane corresponds
to the boundary of the bulk space, and the field $\pi$ to a ``brane-bending" mode.
The two-brane configuration that we have in mind would correspond to a slab of bulk space of finite thickness, with two DGP
branes on its sides. The long-distance physics will be affected by the effective compactification of the extra dimension.
However, we are interested only in the range of scales that are relevant for the Galileon, which remain unaffected if we 
assume that the thickness of the slab of bulk space is sufficiently large. 
The ``brane-bending" modes of the two branes are defined with respect to coordinate systems of opposite orientation
along the extra dimension. If a unique coordinate system is used, one of the two modes must be shifted by the distance
between the two branes and have its sign reversed. At the level of the Galileon theory, 
an equivalent way of describing the two-brane system is by
assuming that the effective theory is given by eq. (\ref{galil}), but with opposite values of the coefficient $M^3_5$ of
the cubic term for each of the two branes. 
 The same assumption must be made for the term (\ref{sn}) of the brane theory, in order to get a 
two-brane configuration symmetric under reflection across the middle plane. 

\section{Solutions of the brane theory}

A class of static solutions of eq. (\ref{eom}) can be obtained if we make the ansatz
$\pi=\pi(w)$ with $w=r^2$.  Eq. (\ref{eom}) becomes
\begin{eqnarray}
\frac{ \lx}{\left(1+4 w \pi_w^2 \right)^{3/2} }\left(3 \pi_w+8 w \pi_w^3+2 w \pi_{ww} \right)
&-&\frac{4M^3_5 \pi_w}{\left(1+4 w \pi_w^2 \right)^{2}}\left( 3 \pi_w+4 w \pi_w^3+4 w \pi_{ww}\right) 
\nonumber \\
&-&\frac{12 M^2_4 \pi_w^2}{\left(1+4 w \pi_w^2 \right)^{5/2}}\left( \pi_w+2 w \pi_{ww}\right)=0,
\label{eoms} \end{eqnarray}
where the subscripts denote differentiation with respect to $w$. 

\begin{figure}[t]
\begin{center}
\epsfig{file=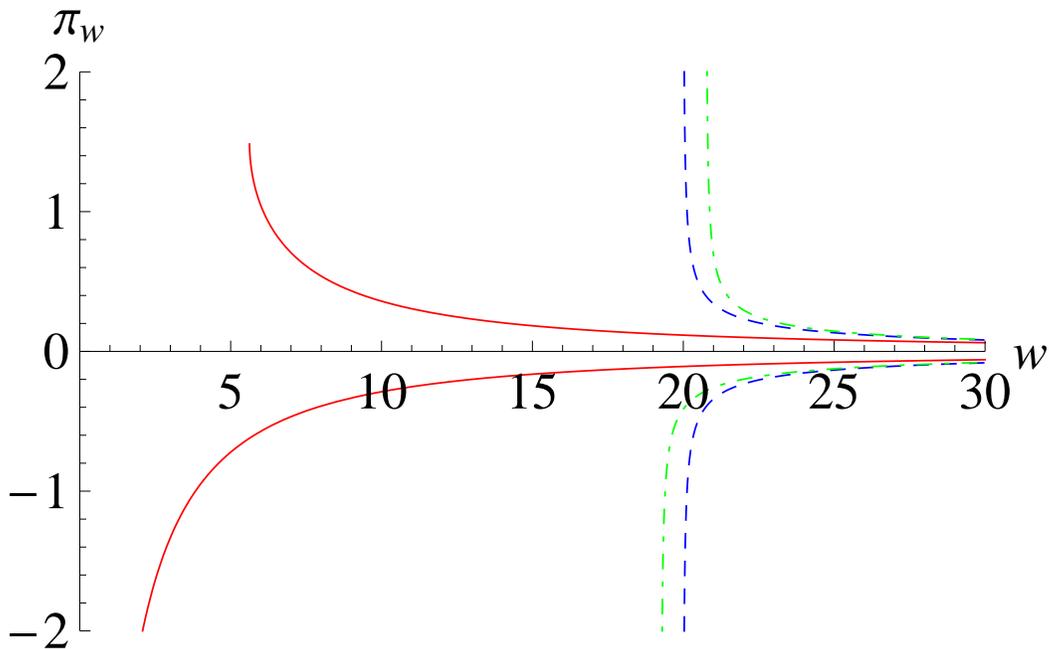,height=9cm}
\end{center}
\caption{The solutions (\ref{sols1}) (dashed), (\ref{sols12}) (dot-dashed), (\ref{solgsta}) (solid) for $\kappa=1$, $c=10$.}
\label{fig2}
\end{figure}

\begin{figure}[t]
\begin{center}
\epsfig{file=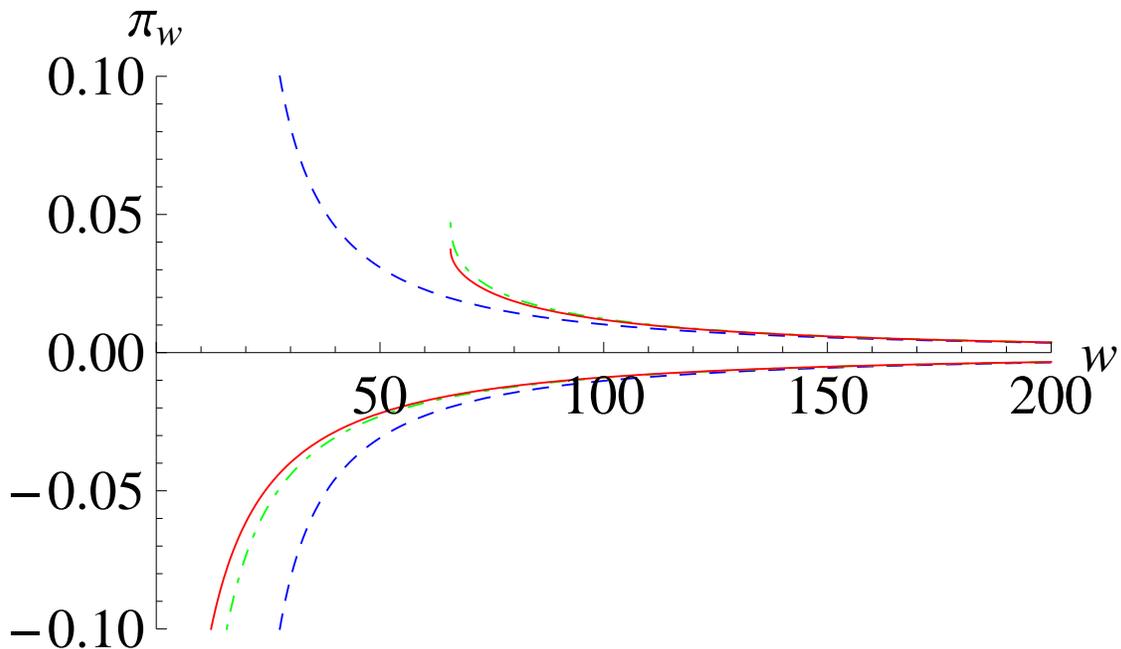,height=9cm}
\end{center}
\caption{The solutions (\ref{sols1}) (dashed), (\ref{sols12}) (dot-dashed), (\ref{solgsta}) (solid) for $\kappa=40$, $c=10$.}
\label{fig3}
\end{figure}

For $M_5=M_4=0$ the action is reduced to the DBI one. 
In this case, the solution of the above equation is
\be
\pi_w=\pm \frac{c}{\sqrt{w^3-4 c^2 w}},
\label{sols1} \ee
with $c>0$. 
The two branches are depicted as dashed lines in figs. \ref{fig2}  and \ref{fig3} for $c=10$. 
Integrating $\pi_w$ with respect to $w$ and joining the solutions 
smoothly at the location of the square-root singularity of eq. (\ref{sols1}) generates 
a continuous double-valued function of $r$, which extends from infinite $r$ to $r_{th}=\sqrt{2c}$ and back to infinity. This catenoidal 
solution describes a pair of branes connected by
a static throat \cite{callan,gibbons}. 
The integration constant $c$ is related to the total energy of the throat $E_{th}$. We have $E_{th} \sim c^{3/4}$, 
with $E_{th}$ and $c$ expressed in terms of $\Lx$ \cite{rizos}. 
The applicability of the effective brane theory for the description of the throat configuration requires $c \gg 1$ in the same units.

The full equation (\ref{eoms}) has similar solutions for nonvanishing $M_5,M_4$. The general solution cannot be 
put in a simple analytical form. However, this is possible for special cases.  
For $\lx=M_4=0$ the solution is
\be
\pi_w=\pm \frac{c}{\sqrt{w^{3/2}-4 c^2 w}},
\label{sols2} \ee
while for $\lx=M_5=0$ it is 
\be
\pi_w=\pm \frac{c}{\sqrt{w}}.
\label{sols3} \ee
Solutions with square-root singularities also exist for nonzero values of more that one of $\lx$, $M_5$, $M_4$.
For $M_4=0$ and $\kx=12 M^3_5/\lx$ we find
\be
\pi_w= \frac{\pm\sqrt{6}c }{\sqrt{3w^3+\sqrt{9w^6\mp12\kx c\, w^{9/2}}-24c^2 w \mp2\kx c \,w^{3/2}}}.
\label{sols12} \ee
This solution is reduced to eq. (\ref{sols1}) for $\kx=0$. 

The solution (\ref{sols12}) describes the modification of the throat in the presence of the higher-order term (\ref{sn}). 
The two branches of the solution (\ref{sols3}) are depicted as dot-dashed lines in fig. \ref{fig2} for $\kx=1$, $c=10$, 
and  in fig. \ref{fig3} for $\kx=40$, $c=10$. 
It is apparent that the location of the singularity is shifted in opposite directions for each of the two branches. The reason can
be found in the form of the term (\ref{sn}), whose sign depends on the sign of the trace $K$ of the extrinsic curvature.
As the latter is proportional to the second derivatives of the field $\pi$, the two branches have values of $K$ of opposite sign. 
We discussed this point at the end of the previous section, where we argued that in a two-brane configuration 
the brane actions must be characterized by opposite values of $M^3_5$. 
It can be checked easily that reversing the sign of $\kx$ in eq. (\ref{sols12}) generates the reflections across the horizontal
axis of the two solutions depicted as dot-dashed lines in figs. (\ref{fig2}), (\ref{fig3}).  
In this way, two solutions of opposite $M^3_5$ can be joined smoothly at the common location of their
singularities, for the construction of a catenoidal configuration.

In the following we shall focus on the branch corresponding to the lower signs in eq. (\ref{sols12}), which allows us to make
contact with the solution of the Galileon theory that realizes the Vainshtein mechanism. 
For large $w$ the solution
falls off $\sim w^{-3/2}$, similarly to eq. (\ref{sols1}). On the other hand, the inner part of the solution is affected by
the presence of the higher-order term (\ref{sn}) in the effective action. The modification of the solution is small for large $c$, as 
can be seen in fig. \ref{fig2}. The singularity of this
branch is located at $w_{th}=2c+\kx(\kx-\sqrt{288c+\kx^2})/72$. For $c\to\infty$ we have $w_{th}\simeq 2c$, while
for $c\to 0$, we have $w_{th} \simeq 144c^2/\kx^2$.

A class of time-dependent solutions of eq. (\ref{eom}) can be obtained through the ansatz $\pi=\pi(z)$, with $z=r^2-t^2$. 
The field equation becomes
\begin{eqnarray}
\frac{ \lx}{\left(1+4 z \pi_z^2 \right)^{3/2} }\left(2 \pi_z+6 z \pi_z^3+z \pi_{zz} \right)
&-&\frac{12M^3_5 \pi_z}{\left(1+4 z \pi_z^2 \right)^{2}}\left( \pi_z+2 z \pi_z^3+z \pi_{zz}\right) 
\nonumber \\
&-&\frac{12M^2_4 \pi_z^2}{\left(1+4 z \pi_z^2 \right)^{5/2}}\left( 2 \pi_z+2 z \pi_z^3 +3z \pi_{zz}\right)=0,
\label{eomd} \end{eqnarray}
where the subscripts denote differentiation with respect to $z$.

\begin{figure}[t]
\begin{center}
\epsfig{file=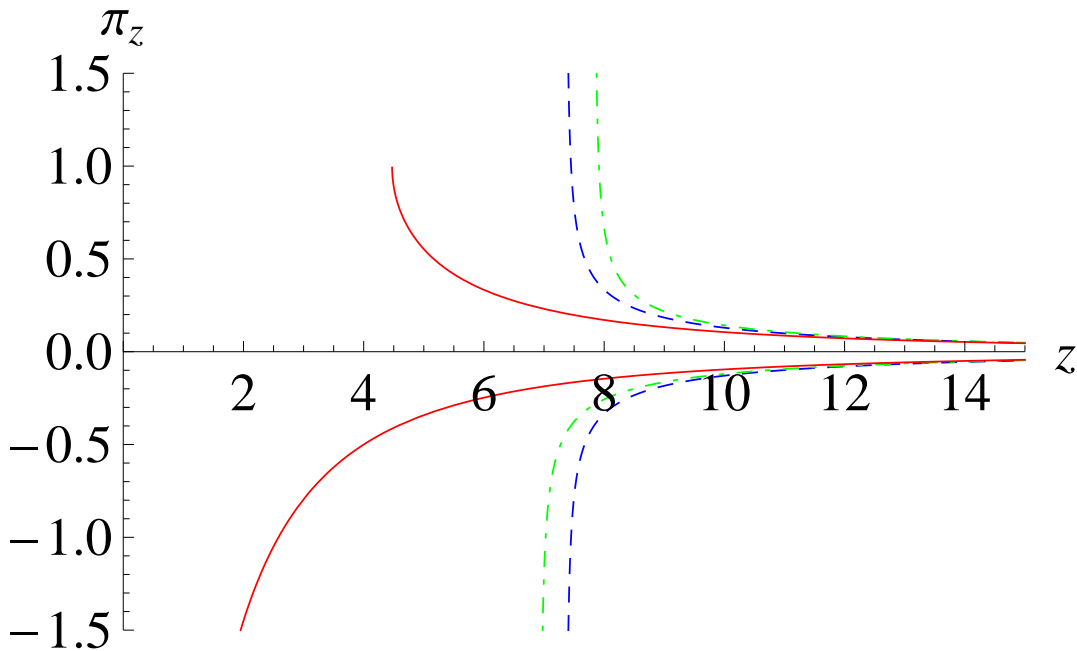,height=9cm}
\end{center}
\caption{The solutions (\ref{sold1}) (dashed), (\ref{sold12}) (dot-dashed), (\ref{solb1}) (solid) for $\kappa=1$, $c=10$.}
\label{fig1}
\end{figure}

For $M_5=M_4=0$ the solution is \cite{rizos}
\be
\pi_z=\pm \frac{c}{\sqrt{z^4-4 c^2 z}},
\label{sold1} \ee
with $c>0$. 
The two branches are depicted as dashed lines in fig. \ref{fig1} for $c=10$. They 
possess square-root singularities at $z_{th}=(2c)^{2/3}$.
Integrating $\pi_z$ with respect to $z$ we obtain 
two curves that can be joined smoothly at $z_{th}$. 
The resulting catenoidal solution $\pi(z)$ describes a pair of branes connected by
a throat of radius $r_{th}=\sqrt{z_{th}+t^2}$. The throat starts with infinite radius for $t\to-\infty$, evolves down
to a minimal size at $t=0$ and subsequently re-expands for positive $t$.

The full equation (\ref{eoms}) has similar solutions for nonvanishing $M_5,M_4$. For $\lx=M_4=0$ the solution is given by 
\be
\pi_z=\pm \frac{c}{\sqrt{z^{2}-4 c^2 z}},
\label{sold2} \ee
while for $\lx=M_5=0$ by
\be
\pi_z=\pm \frac{c}{\sqrt{z^{4/3}-4 c^2 z}}.
\label{sold3} \ee
Similarly to the static case, 
solutions with square-root singularities also exist for nonzero values of more that one of $\lx$, $M_5$, $M_4$.
For $M_4=0$ and $\kx=12 M^3_5/\lx$ we find
\be
\pi_z=\frac{\pm\sqrt{2}c }{\sqrt{z^4+z^3\sqrt{z^2\mp2\kx c}-8c^2 z \mp\kx c \,z^{2}}}.
\label{sold12} \ee
This solution reproduces eq. (\ref{sold1}) for $\kx=0$.
The two branches are depicted as dot-dashed lines in fig. \ref{fig1} for $\kx=1$, $c=10$. 
The construction of a catenoidal solution can be carried out in a way analogous to the static case.
For large $z$ the solution (\ref{sold12})
falls off $\sim z^{-2}$, similarly to eq. (\ref{sold1}). On the other hand, the inner part is affected by
the presence of the higher-order term (\ref{sn}).

\section{Solutions of the Galileon theory}

One of the most interesting features of the Galileon theory is revealed if we consider static, spherically symmetric
solutions of the form $\pi=\pi(w)$, with $w=r^2$.  The equation of motion (\ref{galil}) becomes
\be
\lx\left(3 \pi_w+2w \pi_{ww} \right)
-4M^3_5 \pi_w\left(3 \pi_w+4w \pi_{ww}\right) 
-12 M^2_4 \pi_w^2    \left(  \pi_w+2w \pi_{ww}\right)=0.
\label{eomgstat} \ee
For  $M_4=0$, $\kx=12 M^3_5/\lx$ the solution is 
\be
\pi_w=\frac{3}{2\kx}\left(1-\sqrt{1\mp\frac{4}{3} \frac{\kx c}{w^{3/2}}} \right).
\label{solgsta} \ee
The two branches are depicted as solid lines in  fig. \ref{fig2} for $\kx=1$, $c=10$, and  in fig. \ref{fig3} for 
$\kx=40$, $c=10$. 
The branch with the upper sign terminates at the point $w=(4\kx c/3)^{2/3}$. 
The branch with the lower sign  displays the characteristic form associated with the Vainshtein mechanism. 
For $w\gg w_V$, with $w_V\sim (\kx c)^{2/3}$ the square of the Vainshtein radius, the solution is $\pi_w\sim c \,w^{-3/2}$,  so that 
$\pi\sim c/r$. On the other hand, for $w\ll w_V$, we have $\pi_w\sim \sqrt{c/\kx}\, w^{-3/4}$,  so that 
$\pi\sim \sqrt{c/\kx} \,\sqrt{r}$. This solution requires the presence of a large
point-like source at the origin with mass $\sim c$. The form of the solution at distances smaller than the Vainshtein
radius results in the effective decoupling of fluctuations of the field $\pi$ induced by 
smaller energy sources \cite{vainshtein}. 
The Galileon theory is
applicable at length scales larger than 1 in units of the scale $\Lx$. This means that the Vainshtein mechanism is
operational only for large values of $\kx c$ in the same units. 

The correspondence between the brane solutions  (\ref{sols1}), (\ref{sols12}) and the
solution (\ref{solgsta}) of the Galileon theory is apparent in figs. \ref{fig2} and \ref{fig3}. All solutions involve an 
integration constant and have the same asymptotic form at large distances. 
However, they deviate near the singularity of the brane solutions.
A more precise correspondence can be found 
if we eliminate the term $\sim c^2$ in the denominator of eq. (\ref{sols12}). 
The remaining terms can be rewritten as eq. (\ref{solgsta}). 
We thus conclude that the static solution of the Galileon theory can be obtained from the solution 
(\ref{sols12})  of the brane theory in the formal limits $\kx c\to \infty$ with $c$ held fixed, or $c\to 0$ with $\kx c$ fixed. 
This formal correspondence does not constraint $c$ to be small in units of $\Lx$. As
is apparent from fig. \ref{fig3},  for $\kx$ sufficiently larger than $c$, the brane and Galileon solutions 
almost coincide (apart from the location of the singularity) even for large $c$. For the theory of fig. \ref{fig3} 
the Vainshtein radius is $w_V\simeq 66$. It is clear that the Vainshtein mechanism is operational within the
brane picture as well, for sufficiently large $\kx$.
If we require $c\gg 1$ in units of $\Lx$, the Vainshtein mechanism operates if $\kx \gg c \gg 1$. 
It must be kept in mind, however, that large values of $\kx$ indicate that the trilinear coupling of the
$\pi$-field is large, so that quantum corrections are enhanced. The analysis of quantum corrections goes
beyond the scope of this work. They have been discussed in ref. \cite{quantum} for the Galileon theory and in
ref. \cite{codello} for the brane theory. 

Similarly to our analysis of the 
brane theory, we can obtain time-dependent solutions through the ansatz $\pi=\pi(z)$, with $z=r^2-t^2$.  
The field equation becomes
\be
\lx\left(2 \pi_z+z \pi_{zz} \right)
-12M^3_5 \pi_z\left( \pi_z+z \pi_{zz}\right) 
-12M^2_4 \pi_z^2    \left( 2 \pi_z+3z \pi_{zz}\right)=0,
\label{eomdnr} \ee
which should be compared with eq. (\ref{eomd}) in the brane theory. We can obtain the solution analogous 
to eq. (\ref{sold12})
by setting $M_4=0$ and defining $\kx=12 M^3_5/\lx$.  
If we require that $\pi_z$ vanishes for $z\to \infty$, the solution is
\be
\pi_z=\frac{1}{\kx}\left(1-\sqrt{1\mp 2\frac{\kx c}{z^2}} \right),
\label{solb1} \ee
with $c>0$.
Its asymptotic form for large $z$ matches that of eqs. (\ref{sold1}) and (\ref{sold12}). On the other hand,
the two branches, displayed as solid lines in fig. \ref{fig1} for $\kx=1$, $c=10$,
 have completely different form for small $z$. The branch with the lower sign has a nonintegrable
divergence at $z=0$, while the branch with the upper sign ceases to exist below $z=\sqrt{2\kx c}$. 
It is difficult to give a physical interpretation to either case. For larger values of $\kx$, the solutions (\ref{sold12}) and (\ref{solb1}) 
coincide for most of the range of $z$, apart from the region of the singularities. This behavior is similar to the one
in the static case, depicted in fig. \ref{fig3}. The solutions (\ref{sold12}), (\ref{solb1}) become identical in
the formal limits $\kx c\to \infty$ with $c$ held fixed, or $c\to 0$ with $\kx c$ fixed. 

\section{Discussion}

In the context of the brane theory, the static solution can be interpreted as some type of sphaleron, with 
energy equal to the height of the barrier that must be overcome for the annihilation of the two-brane system
\cite{callan,rizos}. The time-dependent solution describes the dynamics of the annihilation process. 
Through analytical continuation to Euclidean space, this solution can be identified with the corresponding instanton.  
It must be kept in mind, however, that 
the spontaneous nucleation of the throat configuration 
may not be relevant for the annihilation of branes if these 
are identified with the D-branes of string theory. 
The annihilation of D-branes can take place through string processes at a rate faster than the one 
associated with the nucleation of a throat and its subsequent growth \cite{callan}. 

The form of the static and time-dependent solutions within the Galileon theory, eqs. (\ref{solgsta}) and (\ref{solb1})
respectively, 
indicates that they are not appropriate for the description of the throat-nucleation process. The Galileon theory 
is an approximation to the DGP model which is expected to be valid at length scales larger than the inverse of 
the strong-coupling scale $\Lx$. It is appropriate for the description of macroscopic effects,
such as the decoupling of the scalar mode $\pi$ through the Vainshtein mechanism. However, it fails in regions
in which a field configuration has large derivatives, such as near singularities. The embedding of the Galileon theory
within a more complete framework, such as the brane theory, can resolve such problems. 
However, the brane theory cannot be viewed as the UV completion of the Galileon before its quantum
properties are understood.

We note that it is possible to cut the two branches of all the solutions we discussed at a sufficiently large
value $w_b$ or $z_b$ and join
the outer segments. The resulting brane configuration would possess a kink at the location of the throat.
This construction requires explicit boundary terms at $w_b$ or $z_b$, whose presence does not have an
obvious justification.

The derivation of the solutions we presented
was possible because of the special structure of the action of eqs. (\ref{sl})-(\ref{skb}):
Despite its apparent complexity, the field equation of motion (\ref{eom}) does not contain derivatives higher than the 
second. Moreover, the theory involves only derivatives of the field. For the ansatze we employed, which
preserve the rotational or Lorentz symmetry by employing the invariants
$r^2$ and $r^2-t^2$, the equation
of motion becomes a first-order ordinary differential equation, such as (\ref{eoms}) or (\ref{eomd}), which 
can often be solved in closed form. 
This observation explains why theories with kinetic gravity braiding \cite{braiding}, as well as
more general classes of derivative theories included in the generalized Galileon, can have similar solutions \cite{priv}. 
The interpretation of the solutions is not always easy, as in general they have singularities, while the theories do not
have a geometric origin. 
One possibility is to identify the time-dependent solutions with propagating shock fronts, as in refs. \cite{heisenberg,dynamical,rizos}.
The physics of such fronts depends sensitively on the details of the theory.

\section*{Acknowledgments}
I would like to thank A. Vikman for useful discussions and collaboration during the initial stages of this work.
This research has been supported in part by
the ITN network ``UNILHC'' (PITN-GA-2009-237920).
It has also been co-financed by the European Union (European Social Fund - ESF) and Greek national 
funds through the Operational Program ``Education and Lifelong Learning" of the National Strategic Reference 
Framework (NSRF) - Research Funding Program: ``THALIS. Investing in the society of knowledge through the 
European Social Fund".


\begin{thebibliography}{999}



\bibitem{callan}
  C.~G.~Callan and J.~M.~Maldacena,
Nucl.\ Phys.\ B {\bf 513} (1998) 198  [hep-th/9708147].  

\bibitem{gibbons}
  G.~W.~Gibbons,
  Nucl.\ Phys.\  B {\bf 514} (1998) 603
  [arXiv:hep-th/9709027];
[arxiv:hep-th/9801106].  

\bibitem{dynamical}
  J.~Rizos and N.~Tetradis,
JHEP {\bf 1204} (2012) 110  [arXiv:1112.5546 [hep-th]].

\bibitem{rizos}
  J.~Rizos and N.~Tetradis,
JHEP {\bf 1302} (2013) 112  [arXiv:1210.4730 [hep-th]].  

\bibitem{heisenberg}
  W. Heisenberg,
  Zeit.\ Phys.\  Bd. {\bf 133, 3} (1952) 65.


\bibitem{dbigal}
  C.~de Rham and A.~J.~Tolley,
JCAP {\bf 1005} (2010) 015  [arXiv:1003.5917 [hep-th]].  

\bibitem{gliozzi}
 O.~Aharony and M.~Field,
JHEP {\bf 1101} (2011) 065  [arXiv:1008.2636 [hep-th]];  
\\
  O.~Aharony and M.~Dodelson,
JHEP {\bf 1202} (2012) 008  [arXiv:1111.5758 [hep-th]];  
\\
  F.~Gliozzi and M.~Meineri,
JHEP {\bf 1208} (2012) 056  [arXiv:1207.2912 [hep-th]].  

 
\bibitem{codello}
  A.~Codello, N.~Tetradis and O.~Zanusso,
JHEP {\bf 1304} (2013) 036  [arXiv:1212.4073 [hep-th]].  

\bibitem{horndeski}
  G.~Horndeski,
Int.\ J.\ Theor.\ Phys. {\bf 10} (1974) 363.


\bibitem{genegal}
C.~Deffayet, G.~Esposito-Farese and A.~Vikman,
Phys.\ Rev.\ D {\bf 79} (2009) 084003  [arXiv:0901.1314 [hep-th]];
\\
 C.~Deffayet, X.~Gao, D.~A.~Steer and G.~Zahariade,
Phys.\ Rev.\ D {\bf 84} (2011) 064039  [arXiv:1103.3260 [hep-th]];
\\
  T.~Kobayashi, M.~Yamaguchi and J.~'i.~Yokoyama,
Prog.\ Theor.\ Phys.\  {\bf 126} (2011) 511  [arXiv:1105.5723 [hep-th]];
\\
 C.~Charmousis, E.~J.~Copeland, A.~Padilla and P.~M.~Saffin,
Phys.\ Rev.\ Lett.\  {\bf 108} (2012) 051101  [arXiv:1106.2000 [hep-th]].  

\bibitem{galileon}
  A.~Nicolis, R.~Rattazzi and E.~Trincherini,
Phys.\ Rev.\ D {\bf 79} (2009) 064036  [arXiv:0811.2197 [hep-th]].  

\bibitem{dgp}
  G.~R.~Dvali, G.~Gabadadze and M.~Porrati,
Phys.\ Lett.\ B {\bf 485} (2000) 208  [hep-th/0005016];
\\
 C.~Deffayet, G.~R.~Dvali and G.~Gabadadze,
Phys.\ Rev.\ D {\bf 65} (2002) 044023  [astro-ph/0105068].  

\bibitem{vainshtein}
  A.~I.~Vainshtein,
 Phys.\ Lett.\ B {\bf 39} (1972) 393;
\\ 
 E.~Babichev and C.~Deffayet,
arXiv:1304.7240 [gr-qc].  


\bibitem{quantum}
 A.~Nicolis and R.~Rattazzi,
 JHEP {\bf 0406} (2004) 059  [hep-th/0404159];
\\
 K.~Hinterbichler, M.~Trodden and D.~Wesley,
Phys.\ Rev.\ D {\bf 82} (2010) 124018  [arXiv:1008.1305 [hep-th]];
\\
  C.~de Rham, G.~Gabadadze, L.~Heisenberg and D.~Pirtskhalava,
Phys.\ Rev.\ D {\bf 87} (2013) 085017 
[arXiv:1212.4128 [hep-th]].

\bibitem{braiding}
  C.~Deffayet, O.~Pujolas, I.~Sawicki and A.~Vikman,
JCAP {\bf 1010} (2010) 026  [arXiv:1008.0048 [hep-th]];
\\
  O.~Pujolas, I.~Sawicki and A.~Vikman,
JHEP {\bf 1111} (2011) 156  [arXiv:1103.5360 [hep-th]].  

\bibitem{priv}
A.~Vikman, private communication.

\end{thebibliography}
\end{document}